\documentclass[aps,prl,twocolumn,superscriptaddress,longbibliography,amsmath,amssymb]{revtex4-1}

\usepackage{xcolor}
\usepackage{graphicx}

\begin{document}

\title{Radiative lifetime of free excitons in hexagonal boron nitride}

\author{S\'ebastien Roux}
\affiliation{Universit\'e Paris-Saclay, UVSQ, CNRS,  GEMaC, 78000, Versailles, France}
\affiliation{Laboratoire d'Etude des Microstructures, ONERA-CNRS, Universit\'e Paris-Saclay, BP 72, 92322 Ch\^atillon Cedex, France}
\author{Christophe Arnold}
\affiliation{Universit\'e Paris-Saclay, UVSQ, CNRS,  GEMaC, 78000, Versailles, France}
\author{Fulvio Paleari}
\affiliation{CNR-ISM, Division of Ultrafast Processes in Materials (FLASHit), Area della Ricerca di Roma 1, Via Salaria Km 29.3, I-00016 Monterotondo, Scalo, Italy}
\author{Lorenzo Sponza}
\affiliation{Laboratoire d'Etude des Microstructures, ONERA-CNRS, Universit\'e Paris-Saclay, BP 72, 92322 Ch\^atillon Cedex, France}
\author{Eli Janzen}
\affiliation{ Tim Taylor Department of Chemical Engineering, Kansas State University Manhattan, KS 66506, USA}
\author{James H. Edgar}
\affiliation{ Tim Taylor Department of Chemical Engineering, Kansas State University Manhattan, KS 66506, USA}
\author{B\'erang\`ere Toury}
\affiliation{Laboratoire des Multimat\'eriaux et Interfaces, UMR CNRS 5615, Univ Lyon \\ Universit\'e Claude Bernard Lyon 1, F-69622 Villeurbanne, France}
\author{Catherine Journet}
\affiliation{Laboratoire des Multimat\'eriaux et Interfaces, UMR CNRS 5615, Univ Lyon \\ Universit\'e Claude Bernard Lyon 1, F-69622 Villeurbanne, France}
\author{Vincent Garnier}
\affiliation{Laboratoire MATEIS, UMR CNRS 5510, Univ Lyon, INSA Lyon,  F-69621 Villeurbanne, France}
\author{Philippe Steyer}
\affiliation{Laboratoire MATEIS, UMR CNRS 5510, Univ Lyon, INSA Lyon,  F-69621 Villeurbanne, France}
\author{Takashi Taniguchi}
\affiliation{International Center for Materials Nanoarchitectonics, National Institute for Materials Science,  1-1 Namiki, Tsukuba 305-0044, Japan}
\author{Kenji Watanabe}
\affiliation{Research Center for Functional Materials, National Institute for Materials Science, 1-1 Namiki, Tsukuba 305-0044, Japan}
\author{Fran\c cois Ducastelle}
\affiliation{Laboratoire d'Etude des Microstructures, ONERA-CNRS, Universit\'e Paris-Saclay, BP 72, 92322 Ch\^atillon Cedex, France}
\author{Annick Loiseau}
\email{annick.loiseau@onera.fr}
\affiliation{Laboratoire d'Etude des Microstructures, ONERA-CNRS, Universit\'e Paris-Saclay, BP 72, 92322 Ch\^atillon Cedex, France}
\author{Julien Barjon}
\email{julien.barjon@uvsq.fr}
\affiliation{Universit\'e Paris-Saclay, UVSQ, CNRS,  GEMaC, 78000, Versailles, France}

\date{\today}

\begin{abstract}
Using a new time-resolved cathodoluminescence system  dedicated to the UV spectral range, we present a first estimate of the radiative lifetime of free excitons in hBN at room temperature. This is carried out from a single experiment giving both the absolute luminescence intensity under continuous excitation and the decay time of free excitons in the time domain. The radiative lifetime of indirect excitons in hBN is equal to 27 ns, which is much shorter than in other indirect bandgap semiconductors. This is explained by the close proximity of the electron and the hole in the exciton complex, and also by the small energy difference between indirect and direct excitons. The unusually high luminescence efficiency of hBN for an indirect bandgap is therefore semi-quantitatively understood.
\end{abstract}

\maketitle

Hexagonal boron nitride (hBN) is a wide indirect band gap semiconductor \cite{Schue2019} with a honeycomb lattice similar to that of graphene.  It  plays a key role in the world of  2D materials, as it is the best insulating material for use as a substrate or as capping layers of other 2D materials such as transition metal dichalcogenides (TMD) and graphene \cite{Dean2010,Cadiz2017,Ajayi2017}. It also exhibits many promising optical properties in the UV range as well as in the infra-red range: it is a natural hyperbolic material. Some defects in hBN are  good single-photon emitters \cite{Caldwell2019, Fournier2020}. The BN material quality used for these purposes is crucial and the best and most widely used hBN layers by the world community are obtained  by mechanical exfoliation from single crystals grown at high pressure and high temperature  \cite{Watanabe2004}. As this process produces crystals limited in size, alternative crystal growth methods have been proposed such as atmospheric pressure and high temperature (APHT) processes \cite{Kubota2007,Liu2018,Li2020bis} polymer derived ceramics (PDC) processes combined with a sintering process  \cite{Li2018,Li2020}  as well as direct synthesis routes to thin films \cite{Ismach2012,Kim2015,Prevost2020}. Optical spectroscopies of the luminescence properties of hBN are useful   \cite{Schue2016,Sonntag2020} for  assessing the quality of hBN samples. Thus,  for practical and fundamental reasons, it is of the utmost importance to understand the optical properties of hBN.

Indirect optical transitions involving phonon emissions are generally less efficient than direct ones. hBN is an apparent exception to this rule.  First, the fine structure of the upper  luminescence band at 215 nm is linked to phonon-assisted indirect exciton recombination \cite{Cassabois2016,Paleari2019,Canuccia2019}. Second the internal quantum yield measured for this luminescence is surprisingly high, with values up to 50 \%, matching the level expected for direct transitions \cite{Watanabe2004, Schue2019},  and third there is a strong Stokes shift  between absorption and emission \cite{Museur2011}. The latter point has recently been explained thanks to ab initio calculations \cite{Sponza2018,Schue2019}, revealing that the exciton dispersion is significantly flattened  compared to that of free charge carriers. As a consequence absorption is understood to be governed by direct excitonic transitions and luminescence by indirect transitions.

In this letter, we take it a step further by addressing the root cause of the high quantum luminescence efficiency. To this end, experiments were performed on a time-resolved cathodoluminescence (TRCL) setup that allows us  to  simultaneously measure the internal quantum yield $\eta_{i}$ and the excitonic lifetime $\tau$ of the indirect free exciton in each  bulk hBN crystal. The radiative lifetime is then obtained from the relation $ \tau_r = \tau/\eta_i$. From measurements on different samples with quite different quantum yields, we check this linear relationship, which shows the intrinsic character of the radiative lifetime. The quantum yield which measures the relative contributions of radiative  and  non-radiative channels is related to the purity and crystalline quality of  hBN crystals and can also be used as a valuable indicator of this quality \cite{Schue2016}. On the other hand, the luminescence intensity due to the free exciton population  is proportional to the radiative transition probability from the lowest bright exciton state to the ground state. Its inverse is the  radiative lifetime $\tau_r$. We do find a value for $\tau_r$ much smaller than those of other indirect band gap semiconductors such as silicon and diamond. We explain that  from an analysis of the phonon assisted transition rate, using  standard second order perturbation theory taking into account electron-phonon coupling.

The deep UV spectroscopy spectra were recorded at  room temperature in a JEOL7001F field-emission-gun scanning electron microscope (SEM) coupled to a Horiba Jobin-Yvon cathodoluminescence (CL) detection system as described in Refs \cite{Schue2019} and \cite{Schue2016Nanoscale}. The CL intensity under continuous beam excitation (cw-CL) is measured within an uncertainty of 10\% thanks to a monochromator equipped with a silicon CCD camera. The setup was calibrated with a reference deuterium lamp \cite{Schue2019}, which allows measuring the absolute CL intensity. From the incident power, the instrumental parameters and the material refractive index, it provides an estimate of the luminescence internal quantum yield with an uncertainty of about 50\% \cite{Schue2019}. It indicates the part of the free exciton population that recombines via photon emission.

For the time-resolved CL (TRCL) experiments, a fast-beam blanker was designed, built at the laboratory and installed inside the SEM column. It consists of two metallic plates, polarized by a voltage generator (Avtech - AVNN-1-C-OT. 25 to 200 MHz) to quickly deviate the incident electron beam. The TRCL signal is detected with a customized UV photo-multiplicator (Photonis - MCP-PMT) mounted on the second exit port of the monochromator. The PM operates in single-photon mode with low noise in a 200-550 nm range and its signal is recorded with a counting card (BECKER \& HICKL: SPC-130 EM) synchronised with the voltage generator. The temporal resolution of the TRCL setup is measured at 100 ps. 

The temporal dynamics of the free exciton population was probed following the time-dependent CL intensity, filtered at $ 215 \pm 7.5 $ nm thanks to the monochromator grating. The fast PM was calibrated from the CCD camera so that the internal quantum yield can also be deduced  from the measurements of  the steady-state CL intensity before beam interruption. With this experiment, the exciton decay time and the internal quantum yield are both measured within a single TRCL experiment recorded from a chosen region of interest  at a nanometer scale. In this study, the electron beam excitation is at a high voltage of 15 keV to avoid non-radiative recombinations from the surface \cite{Schue2019} and at a low current of 85 pA to prevent non-linear effects, such as exciton-exciton annihilation \cite{Plaud2019}.

The bulk hBN crystals studied in this work were synthesized using three different growth methods,  as listed in Table \ref{toto}. A high-pressure high-temperature crystal  \cite{Watanabe2004} was analysed, as the hBN reference in the 2D material community (HPHT sample). The second  family of hBN  samples was grown at atmospheric-pressure high-temperature from Ni/Cr or Fe solvents \cite{Kubota2007} (samples APHT\# 1, 2 and APHT \#3 through  \#5 respectively). With this synthesis technique, the boron isotope content can be controled \cite{Liu2017} (samples APHT\#4 and \#5). The last sample was synthesized using an alternative method: the polymer derived ceramics (PDC) route followed by high temperature annealing under moderate pressure(PDC sample)  \cite{Li2020}. The crystals studied are much thicker than the 3~$\mu$m stopping depth of 15 keV electrons in hBN. All  samples were mounted together on the same rack and measured under the same conditions in a single series of TRCL experiments.
\begin{figure}[h]
 \begin{center}
 \includegraphics[scale=0.3]{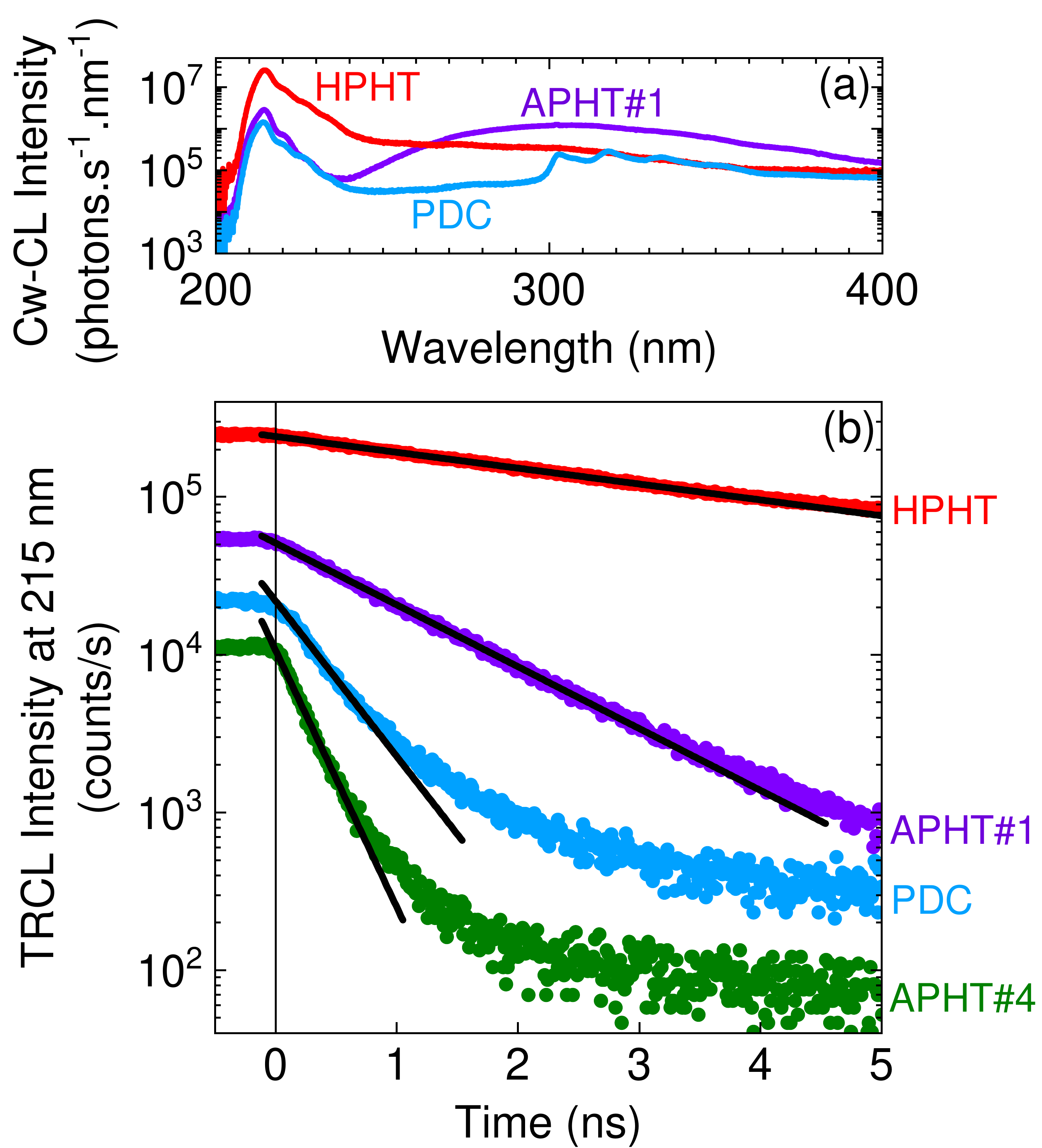}
 \caption{(a) cw-CL spectra at  room temperature of hBN crystals grown from three different methods HPHT, APHT and PDC. The spectra are dominated by the radiative recombination of free excitons  at 215 nm in hBN. (b) Temporal decay of the free exciton luminescence after interruption of the incident electron beam at initial time. The CL intensity under continuous excitation is measured simultaneously before the beam interruption. More information on the samples are given in in Table \ref{toto}.}
 \label{f1}
 \end{center}
\end{figure}
\begin{table}[ht]
\caption{\label{tab:table2}
Fabrication methods and boron isotopic purity  for the samples studied in this work. Results of TRCL experiments for the internal quantum yield ($\eta_{i}$), the exciton lifetime ($\tau$) and the estimated radiative lifetime of the free exciton ($\tau_{r}$=$\tau$/$\eta_{i}$).
}
\begin{ruledtabular}
\begin{tabular}{cccccc}
Sample & Thickn. ($\mu$ m) & Isotopic comp. & $\eta_{i}$ (\%) & $\tau $ (ns) & $\tau_{r} (ns)$ \\
\hline
HPHT & 300 & natural & 18 & 4.2 & 23 \\
APHT\#1 & 100 & $100\% B_{10}$ & 4.2 & 1.12 & 27 \\
APHT\#2 & 5 & natural & 3.5 & 0.99 & 28  \\
PDC & 55 & natural & 1.7 & 0.43 & 25 \\
APHT\#3 & 8 & natural & 1.6 & 0.45 & 28 \\
APHT\#4 & 25 & $100\% \, B_{11} $ & 0.85 & 0.26 & 31 \\
APHT\#5 & 5 &  $50\% \, B_{10} \, \text{and} \, B_{11}$ & 0.14 & $<$ 0.1 & - \\
\end{tabular}
\end{ruledtabular}
\label{toto}
\end{table}

Figure \ref{f1}(a) shows typical cw-CL spectra of hBN crystals made from the HPHT, APHT and PDC routes. The main luminescence feature is  a maximum peak at 215 nm, corresponding to the indirect exciton recombination assisted with optical phonons. Note that all phonon lines involved in the indirect recombinations observed at low temperature \cite{Schue20162Dmaterials} are not resolved at  room temperature due to thermal broadening. The dominant CL peak at 215 nm probably involves both transverse and longitudinal optical phonons at 300K. A residual luminescence from deep defects is also observed at higher wavelengths. In samples grown from the PDC route, color centers are detected with a zero-phonon line at 302 nm, as reported in the literature from carbon-rich regions of HPHT samples \cite{Onodera2019}. In APHT crystals, the colour centers are absent but there is a broad  band extended from 250 to 400 nm also observed in many defective crystals (superimposed or not on the colour centers)\cite{Museur2008, Museur2009} and which will be discussed elsewhere. Figure \ref{f1}(b) shows  the luminescence decay at 215 nm after the electron beam interruption at initial time. The TRCL signal follows a single exponential law over several decades, with a characteristic time corresponding to the free exciton lifetime. This figure directly shows that the cw-CL intensity (before decay) is an increasing function of the exciton lifetime.
\begin{figure}[ht]
 \begin{center}
 \includegraphics[scale=0.35]{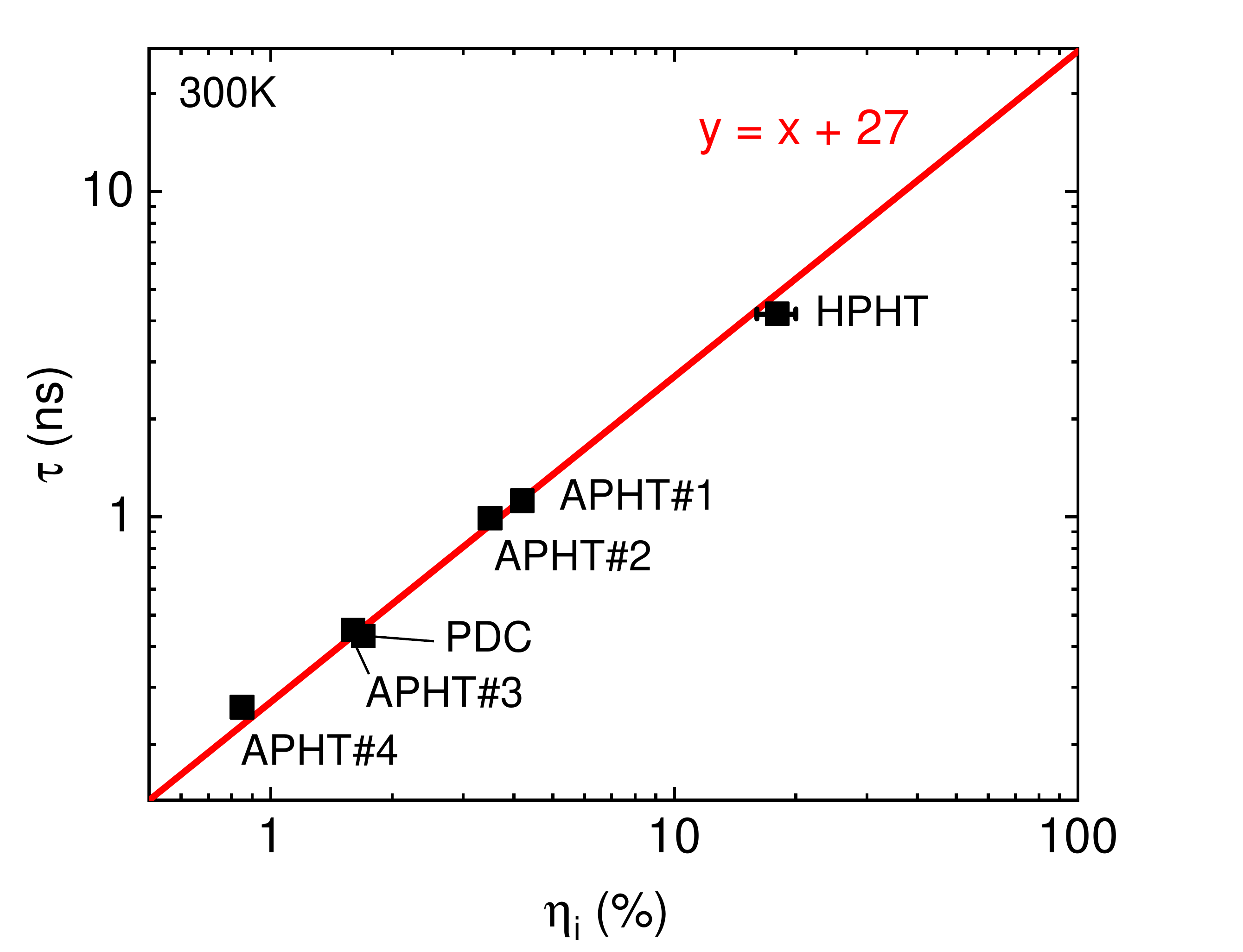}
 \caption{Free exciton lifetime ($\tau$) as a function of the internal quantum yield for the intrinsic luminescence in hBN. The radiative lifetime of hBN indirect excitons is estimated at 27 ns from the linear fit (red curve).}
 \label{f2}
 \end{center}
\end{figure}
In Figure \ref{f2}  the measured lifetimes $ \tau $ of free excitons are plotted as a function of the quantum yields $ \eta_{i} $ measured from the  plateau intensities for the hBN samples described in Table \ref{toto}. The error bars correspond to  two measurements made a few micrometers apart on each sample (see also  Supp. Mat.). 

These results confirm that, as expected,  the internal quantum yield is proportional to the free exciton lifetime, since  $ \eta_{i} = \tau / \tau_{r}$. and that both the free exciton lifetime $\tau$ and its luminescence yield $ \eta_{i} $ are global indicators of the crystal quality. More interestingly, the slope of the linear regression plotted in Figure \ref{f2} corresponds to the radiative recombination lifetime $\tau_{r}$ of free excitons in hBN. We find $\tau_{r} = 27 $ ns. Due to the uncertainty on the measurement of $\eta_{i}$,  the uncertainty on $\tau_{r}$ is typically 50\%.

Many excitonic properties of hBN begin to be understood now \cite{Schue2019,Lorenzo2018,Caldwell2019}.  In the case of a single layer the lowest exciton is doubly degenerate with a very large binding energy (about 2 eV). It has also  very small, on the order of 1 nm. In bulk hBN it gives rise to two doubly degenerate states which can be viewed as bonding and antibonding combinations of localized excitons similar to the single layer 2D state. Due to the symmetry of the AA' stacking of the usual phase of hBN, the lowest bonding state is dark below the bright antibonding state a few tens of meV above. Because of screening effects, the binding energy of the direct exciton is about 700 meV, lower than that of the single layer but its wave function is quite similar and strongly localized in the planes. Actually this applies to the situation corresponding to direct transitions, but, whereas the optical gap of the single layer is direct, the gap of bulk hBN is indirect and the corresponding excitonic state has an energy $E_i$ slightly below $E_d$, the energy of the direct exciton: $E_d -E_i = 70$ meV as measured in \cite{Schue2019}. The indirect exciton has a binding energy about $300$ meV \cite{Schue2019}.

The radiative lifetime of an exciton is proportional to $1/|S|^2$  where $S$ is the associated transition matrix element for the phonon assisted transition. $S$ can be calculated using second order perturbation theory and  in its simplest form is given by \cite{Elliott1957,Toyozawa1958,Toyozawa2003}:
\begin{equation}
S = \frac{M_{dX} M_{ephX}}{E_d-E_i +\hbar\omega_{ph}}  \; ,
\label{perturb}
\end{equation}
where $M_{dX}$ is the usual optical matrix element for the direct transition and $M_{ephX}$ is the electron-phonon matrix element coupling the direct and indirect excitonic wave functions; $\hbar\omega_{ph}$ is the phonon frequency involved in the transition. This analysis has also been used by  Laleyan et al., but without considering explicitly excitonic effects \cite{Laleyan2018,Mengle2019}. More complete calculations based on the Bethe-Salpeter equation or on density matrix methods have been performed recently \cite{Paleari2019,Canuccia2019,Brem2020,Chen2020}.

The value of the simple formula above is to identify the effects responsible for the different behaviours of indirect semiconductors. The first matrix element $M_{dX} $ or more precisely $|M_{dX} |^2 $ is related to the oscillator strength of the direct transition. For an excitonic state it is related to the exciton wave function calculated for zero separation between the electron and the hole. Within a simple Wannier-Mott model the oscillator strength is therefore proportional to $1/a^3_{X}$, where $a_X$ is the Bohr radius of the excitonic state. For an anisotropic system such as hBN, where furthermore lattice effects are important, corrections should be introduced, Nevertheless an approximation can be made  that the oscillator strength strongly increases when the size of the exciton decreases. We compare three indirect semiconductors, Si, diamond and hBN. Within  the Wannier-Mott model, the mean electron-hole distance $\langle d_{eh} \rangle = 3a_X/2$ is equal to 73 \AA {} and 23 \AA, for Si \cite{McLean1960} and diamond \cite{Cardona2001}, respectively. For hBN, the validity of the Wannier-Mott model is limited as shown by the flattening of its exciton dispersion \cite{Schue2019}. $\langle d_{eh} \rangle$ was then calculated from ab initio exciton wavefunctions : it yields 7 \AA {} for the indirect exciton and 4.2 \AA {} for the direct one.

The second factor is the electron-phonon matrix element $M_{ephX}$, which is difficult to estimate without appropriate ab initio calculations. In the case of diamond and hBN the relevant electron-phonon coupling is principally related to the strengths of the C-C or B-N bonds which are similar and larger than that of Si-Si bonds, but since the coupling with excitons is local, the matrix element is also sensitive to the size of the excitons, and should vary qualitatively as the first one $M_{dX}$. Notice here that for a ionic compound such as hBN the most important contributions to the electron-phonon coupling are usually the Fr\"{o}hlich ones but, in the case of neutral excitons, the electron and hole contributions compensate and this is not necessarily  the case \cite{Toyozawa2003,Chernikov2012}.

Finally we have to compare the denominators in Eq.(\ref{perturb}). This is an important point here, because in the case of hBN, the difference $E_d -E_i $ is very weak, $70$ meV  to be compared with values of the order of 2 eV for diamond and silicon. In the case of hBN the main phonon contributions are due to the optical modes, for which $\hbar\omega_{ph} \simeq$ 160 meV (TO mode) and 185 meV (LO mode)
\footnote{In the case of absorption this phonon energy has a minus sign so that the denominator is vanishing. This means that both direct and indirect channels are active together, which requires going beyond perturbation theory.}.

If we compare the different indirect semiconductors, even assuming similar electron-phonon couplings, we obtain that $|S|^2$, proportional to $\tau^{-1}_r$, is a few $10^3$ times larger for hBN than for diamond, and more than $10^4$ times larger than for Si. 

\begin{figure}[ht]
 \begin{center}
 \includegraphics[scale=0.3]{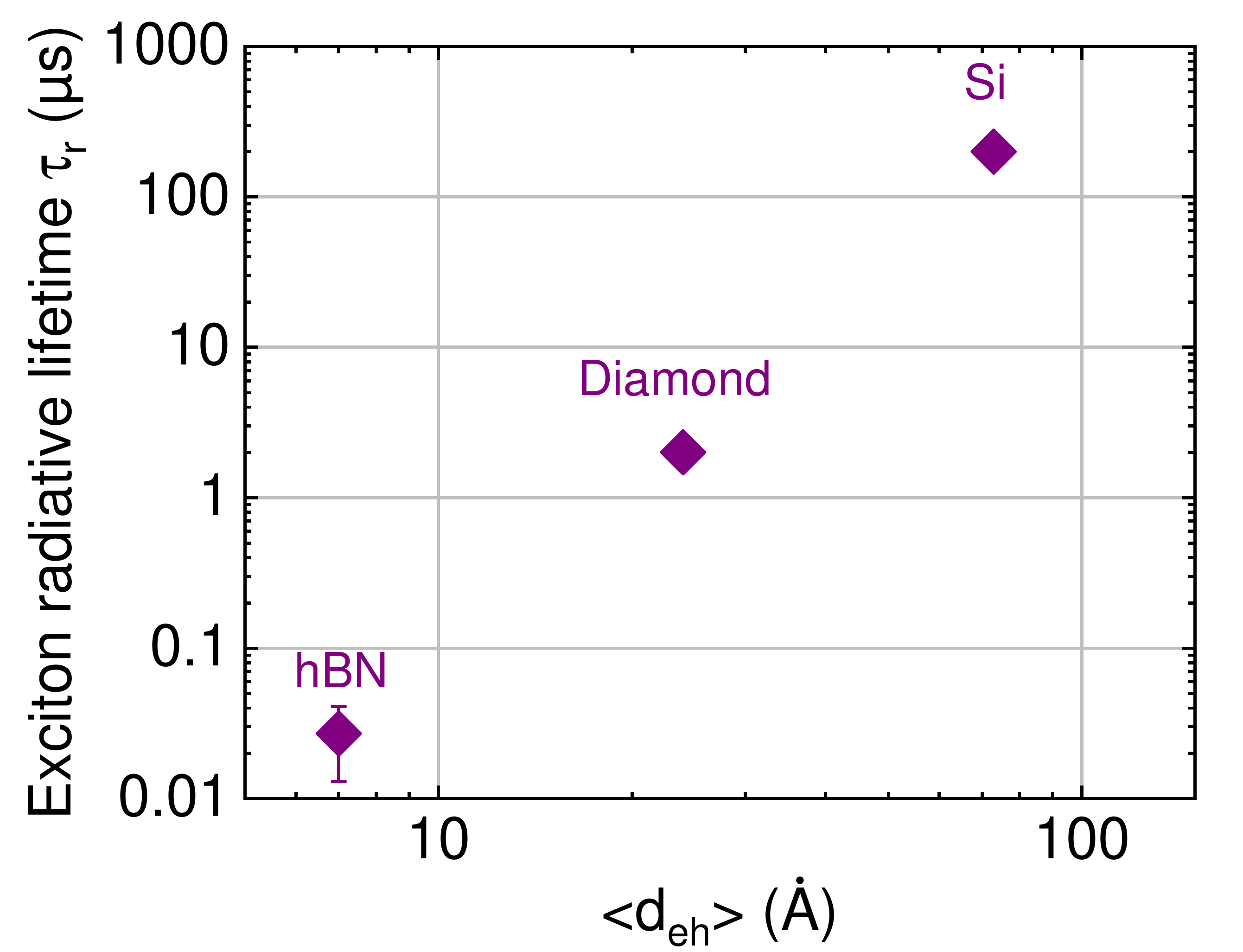}
 \caption{Radiative lifetimes of free excitons in silicon \cite{Cuthbert1970}, diamond \cite{Konishi2020} and of hexagonal boron nitride [this work] as a function of the average distance between electron and hole in  the indirect exciton.}
 \label{f3}
 \end{center}
\end{figure}

Figure \ref{f3}  compares  the radiative lifetimes of these semiconductors as a function of the electron-hole distance $\langle d_{eh} \rangle $. We have taken $\tau_r = 2\, \mu$s and $200\, \mu$s for diamond \cite{Konishi2020}  and Si \cite{Cuthbert1970}, respectively. These lifetimes are 100 times and $10^4$ times larger than for hBN, respectively, to be compared to our value for hBN, about 20 ns. Our theoretical estimate is therefore too large, but several factors can explain that. First there are the experimental uncertainties  mentioned before, but also the fact that  formula (\ref{perturb}) is based on a second order perturbation theory which is no longer valid when the denominator is small, which is the case for hBN. Actually self-energy corrections included in the formalism by Toyozawa and others, could improve the situation, but better ab intio theories are clearly needed. Such preliminary calculations have been performed with a partial treatment of electron-phonon interactions. A secondary conclusion of this work is that the indirect oscillator strength for hBN is about 100 times less than that of the direct transition \cite{Canuccia2019}. This corresponds to a hypothetical radiative lifetime about 200 ps for the direct exciton, which is reasonable \cite{Chen2019}. 

In conclusion, we have developed a new TRCL system dedicated to the UV spectral range which provides a first estimate of the radiative lifetime of free excitons in hBN. The radiative lifetime is evaluated from a single experiment giving both the absolute luminescence intensity under continuous excitation and the decay time of free excitons in the time domain. By comparing hBN crystals issued from different synthesis routes we found that exciton radiative lifetime could be defined intrinsically from the proportionality measured between the free exciton lifetime and the internal quantum yield, and that, with a value equal to  27ns, which is much shorter than in other indirect bandgap  semiconductors. We have performed an analysis of the matrix elements of the phonon assisted transition rate, using  standard second order perturbation theory taking into account electron-phonon coupling. The observed behaviour is explained first by the close proximity of the electron and the hole in the exciton complex, and second by  the weak energy difference between direct and indirect transitions. The unusually high luminescence efficiency of hBN for an indirect bandgap is therefore semi-quantitatively understood.  

Beyond these fundamental aspects, our results have a practical application to compare quantitatively hBN samples. The decay time of free excitons provides an accurate scaling of hBN sample quality, and can be measured from a single experiment carried out at room temperature. This tool can be capitalized for linking  hBN quality to the expected performances of 2D materials in devices using hBN layers.

\begin{acknowledgments}
The research leading to these results has received funding European Union's Horizon 2020 research and innovation program under grant agreements  No 785219 (Graphene Core 2)  and No 881603 (Graphene Core 3). Support for the APHT hBN crystal growth comes from the Office of Naval Research, Award No. N00014-20-1-2474. K.W. and T.T. acknowledge support from the Elemental Strategy Initiative conducted by the MEXT, Japan (Grant Number JPMXP0112101001) and  JSPS KAKENHI (Grant Numbers 19H05790 and JP20H00354).
\end{acknowledgments}

%


\end{document}


\preprint{aps/123-QED}

\onecolumngrid

\section*{Supplementary materials: Radiative lifetime of free excitons in hBN}

\subsection*{Experimental details}
In Figure S1 are plotted the decays of luminescence at 215 nm after the electron beam interruption at t=0 ns. The decays follow a single exponential law over several decades, with a characteristic time associated to the free exciton lifetime. One can observe than the longer the lifetime, the higher the intensity of luminescence under continuous excitation at t < 0 ns. In addition, for some samples, such as the PDC one, a second decay time is observed with a much longer lifetime, of the order of 100 ns. This is why, for measuring $\tau$, we choose to approach the beginning of the decay by an asymptote rather than by a single exponential fit. Also, note that we were not able to measure the excitonic lifetime of the APHT\#5 sample which is shorter than the temporal resolution of our setup. Two measurements were recorded a few micrometers apart on each sample to evaluate their reproducibility. A big deviation of 20\% is observed on APHT\#5 and HPHT samples, it is attributed to homogeneity issues.  For all the other samples, the typical deviation of the $\tau$ and $\eta_{i}$ measurements stays within 5\%, which shows the good homogeneity of the areas analysed, as well as the good reproducibility of $\tau$ and $\eta_{i}$ measurements. 
\begin{figure*}[!h]
 \begin{center}
 \includegraphics[scale=0.60]{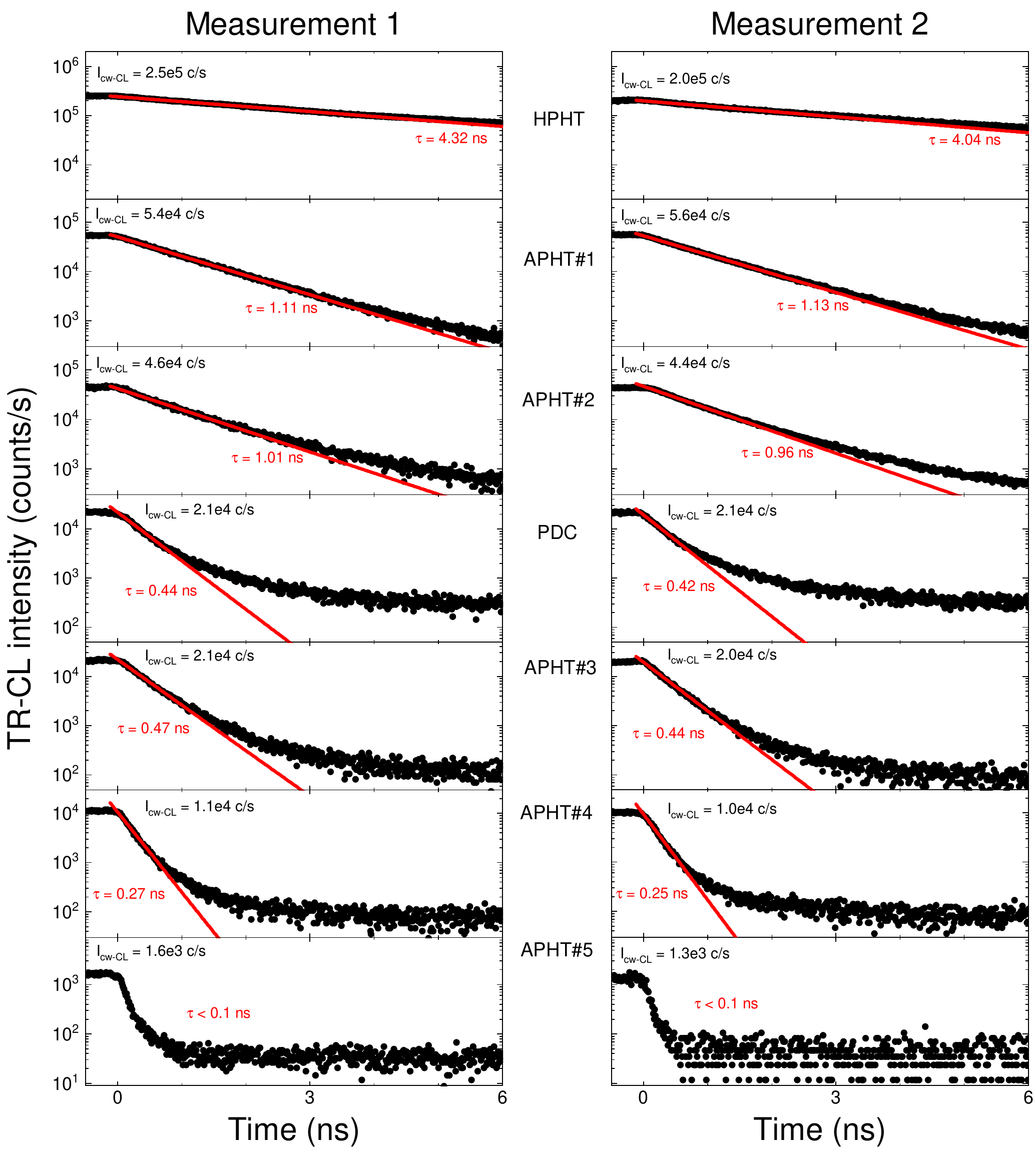}
 \caption{Temporal decays  of luminescence at 215 nm for different samples.}
 \label{FIG. S1}
 \end{center}
\end{figure*}

\newpage

\subsection*{Average electron-hole distance}

The calculation of the average electron-hole distance is based on GW-BSE results obtained by Fulvio Paleari and published in his doctoral thesis~\cite{fulvio_tesi}. 

\bigskip

The DFT calculations where performed in LDA using the free simulation package Quantum Espresso~\cite{quantum-espresso} using FHI pseudopotentials~\cite{pseudos} for the core electrons. The energy cutoff was 110~Ry and the $\mathbf{k}$-point sampling has been done in a $(12 \times 12 \times 4)$ regular mesh. The cell parameters used are $a=4.7177$~Bohr and $c/a=2.6681$.

The GW and Bethe-Salpeter (BSE) calculations were carried out with the free software Yambo~\cite{yambo1,yambo2}. Quasiparticle corrections were applied through a scissor operator of 1.96~eV with a stretching factor of 1.08 in conduction and 1.16 in valence. These values were fixed on the basis of a previous GW calculation~\cite{fulvio_tesi}.
In the BSE calculation, two bands in conduction and two in valence were included to construct the excitonic Hamiltonian, while the electron-hole screening was calculated within RPA including local fields (cutoff energy 10 Ry) and summing transitions over 280 bands.   
These parameters where considered to be at convergence when differences in the excitonic peak position were lower than 0.01 eV. The Coulomb truncation method was used to prevent interactions between replicas of the system and a vacuum of 20~\AA\, was included in the $z$ direction. In order to have a sufficiently fine grid to calculate the excitonic wavefunction at finite momentum, the BSE calculations have been carried out on a $(36 \times 36 \times 6)$   $\mathbf{k}$-point grid.

\bigskip

After diagonalisation of the excitonic Hamiltonian, the excitonic wavefunction $\Psi(\mathbf{r}_e, \mathbf{r}_h)$ is obtained for generic positions of the electron $\mathbf{r}_e$ and the hole $\mathbf{r}_h$ for both the direct and the indirect excitons.
Successively, the hole position $\mathbf{r}^0_h$ has been fixed at 1~\AA\,above a N atom placed at the origin of the Cartesian space (this account for the $p_z$ character of the top valence states). The probability  $P(\mathbf{r}_e) =  \Psi_0(\mathbf{r}_e) |^2$ hence obtained represents the probability finding the bound electron at point $\mathbf{r}_e$. This has been sampled in real space in a regular 3D grid with volume element $d\Omega \approx 0.02$~\AA$^3$ in both calculations and the average electron-hole distance has been calculated by evaluating numerically the integral 
\begin{equation}
\langle d_{eh} \rangle = \int_{\Omega} P(\mathbf{r}) |\mathbf{r} - \mathbf{r}^0_h| d\mathbf{r} \approx \sum_{i} P(\mathbf{r}_i |\mathbf{r}_i - \mathbf{r}^0_h| d\Omega \,,
\end{equation}
where $\Omega$ is the volume of the 3D integration volume.
The result gives $\langle d_{eh} \rangle  = 6.96$~\AA\, in the indirect exciton and $\langle d_{eh} \rangle = 4.16$~\AA\, in the direct one.